\documentclass[twocolumn,showpacs,preprintnumbers,amsmath,amssymb,prl]{revtex4}

\usepackage{amssymb,amsfonts,amsmath}
\usepackage[dvips]{graphicx}
\usepackage[usenames,dvipsnames]{color}
\usepackage{url}
\usepackage[normalem]{ulem}
\usepackage{cancel}

\begin{document}

\preprint{APS}

\title{``Memory foam" approach to unsupervised learning}

% ------------Define authors and affiliations------------------

\author{Natalia~B.~Janson}
\email[E-mail: ]{N.B.Janson@lboro.ac.uk}
\author{Christopher~J.~Marsden}
\affiliation{School of Mathematics, Loughborough University, Loughborough LE11 3TU, UK}

% ----------------Abstract part--------------------------------

\begin{abstract}
We propose an alternative approach to construct an artificial learning system, which naturally learns in an unsupervised manner. 
Its mathematical prototype is a dynamical system, which automatically shapes its vector field in response to the input signal. 
The vector field 
converges to a gradient of a multi-dimensional probability density distribution of the input process, taken with negative sign. The most probable patterns 
are represented by the stable fixed points, 
whose basins of attraction are formed automatically. 
The performance of this system is illustrated with musical signals. 
\end{abstract}


\maketitle

% ---------------- Introduction part -----------------------------

The tasks being posed to, and solved by, the modern
artificial ``intelligent" (AI) devices are broad and include image and speech recognition, machine vision,  language processing and medical diagnostics, to mention just a few \cite{Arbib}.  However, in spite of the word ``intelligence" behind the AI abbreviation, in essence, these machines are  only able to perform two tasks: classification and optimization, which include decision-making.  Learning has been understood merely as acquiring the ability to perform these tasks. 

The performance of modern AI devices is based on algorithms, i.e. while fulfilling their goal they perform a sequence of pre-defined commands. Even the later generation of AI devices, that are based on  neural networks, employ algorithms at least at the stage of learning \cite{Hertz}. 
Contrary to that, it seems that a biological brain does not naturally execute a sequence of commands, although it can be trained to do so (often with some effort,  e.g. when solving routine mathematical problems). In particular, the brain does not seem to {\it learn} by an algorithm.

As can be expected from algorithm-based devices, the natural way of learning generally requires a teacher -- i.e. a {\it truly} intelligent system -- and can be fully supervised, semi-supervised \cite{Semi-Supervised} or reinforcement \cite{Dayan_reinforce}. 
The unsupervised learning defined within the AI field, is acquiring the ability to attribute a new entry to a certain class without any help from a teacher \cite{Dayan_unsupervised}. 

In this Letter we propose an alternative approach to  describe a learning process. Namely, we suggest that a thinking system should work as a machine, that  adjusts its {\it architecture} in response both to sensory input, and to the processes inside itself in an analogue (i.e. non-algorithmic) way. We introduce a mathematical prototype of this machine -- a dynamical system, that shapes its vector field in response to the {\it external} stimulus -- i.e. we describe the first component of the thinking process.  
The model does {\it not} rely on any biological knowledge.  

Let every (scalar or vector) value of the input at the given time moment represent a certain pattern, that can be of any origin: visual, auditory, tactile, olfactory, or their combination. It could be the color of the image, the pitch of the sound, etc. 
The implementation of a non-algorithmic classification (pattern recognition) was proposed in \cite{Hopfield_attractors_86} by means of neural networks (NNs) -- a collection of units, each with fixed architecture, which are flexibly coupled to each other. However, learning in a NN is algorithmic and consists in adjusting the strengths of couplings (``weights")  in response to a training set of patterns. As a result, an energy profile is formed in the phase space of the NN \cite{comment1}, whose minima (attracting fixed points) represent the centres of classes, and the respective basins of attraction represent classes. When learning is over, the weights are fixed, the new input patterns are given by initial conditions, and classification occurs non-algorithmically as the NN evolves towards the nearest attractor \cite{Hertz}. A series of technical problems can occur as a NN learns, including the formation of spurious attractors. Also, the most natural way of learning for a NN is supervised, while semi- or unsupervised learning require considerable complication of the algorithms. 

Here, we propose the construction of a dynamical system, whose vector field is the gradient of the potential energy, which 
is shaped by the external stimulus {\it non-algorithmically} and {\it without supervision}. If the stimulus comes from a stationary and ergodic random process, 
this ``energy" represents a negative multi-dimensional probability density distribution of the input, and  each stable fixed point represents  the most probable pattern from the input class. 
%This way, there are no spurious attractors.  
The system recognizes the new patterns just like a particle that is placed into a potential energy profile $V(x)$, which moves towards the nearest minimum, possibly being affected by noise, according to \cite{Malakhov_97}
\begin{equation}
\label{eq_particle}
\dot{x}=-\frac{\partial V(x,t)}{\partial t} + \xi(t),
\end{equation}
where $x$ represents the location in $N$-dimensional space, and $\xi(t)$ is noise. 

~\vspace{-2mm}

\begin{figure}
%\scalebox{0.8}{\includegraphics{fig_foam_ill.eps}}
\includegraphics[width=0.4\textwidth]{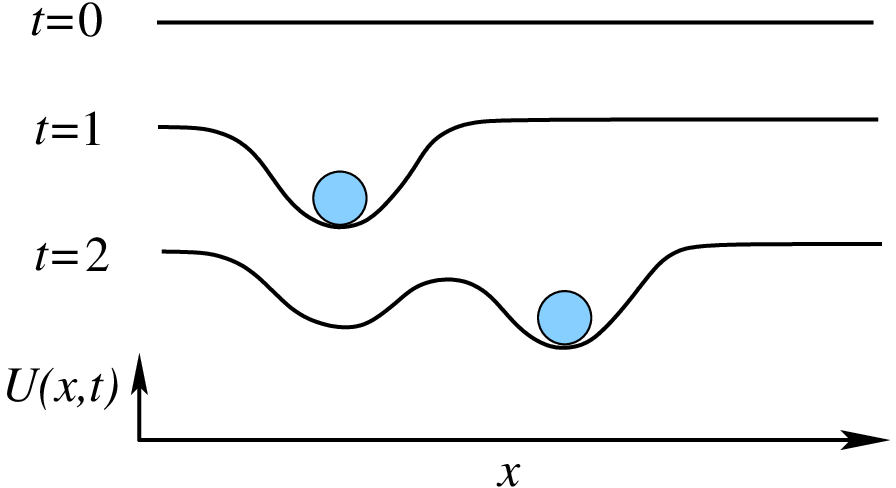}
\caption{\label{fig_foam_ill} (Color online.) Illustration of the idea of memory foam. }
\end{figure}

\noindent {\bf Model.} It is based on a loose analogy with the ``memory foam",  used in orthopedic mattresses, that takes the shape of the body pressed against it, but slowly returns to its original shape after the pressure is removed.  
Assume that initially we have a {\it one-dimensional}  ``foam" stretched in $x$ direction, and that initially it is flat, i.e. its profile is $U(x)$$=$$0$ (Fig. \ref{fig_foam_ill}, $t$$=$$0$). If a stone drops onto the foam at position $x$$=$$\eta$, the foam 
profile is deformed: a dent appears, which is the deepest exactly at
$x$$=$$ \eta$, and gets shallower at larger distances from $\eta$ (Fig. \ref{fig_foam_ill}, $t$$=$$1$). 
Also, assume that the foam is elastic with elasticity factor $k$, that models the capacity to forget. The deeper the dent at the position $x$ is, the faster the foam tries to come back to $U$$=$$0$ (to forget). In other words, the foam will learn about the stone and its position. 
Now assume that we subject the foam to an external stimulus $\eta (t)$, as if at any new time moment $t$ a new stone drops at a new position $x$$=$$\eta(t)$ (Fig. \ref{fig_foam_ill}, $t$$=$$2$), thus shaping the ``foam" continuously.  The signal  $\eta (t)$ can be of either deterministic, or stochastic nature, and can have arbitrary statistical properties.
 Next we derive an equation, that describes the evolution of the foam profile $U(x, t)$ 
under the influence of $\eta (t)$. 

Consider how the foam profile changes over a small, but finite time interval $\Delta t$:
\begin{equation}
\label{eq1}
U (x, t +\Delta t) = U (x, t) - g(x - \eta) \Delta t - k U (x, t) \Delta t, 
   \end{equation}
where $g(z)$ is some non-negative bell-shaped function, describing the shape of a single dent, e.g. a Gaussian function, $g(z)$$=$$\frac{1}{\sqrt{2 \pi \sigma^2_z}} \exp ( - \frac{z^2}{\sigma^2_z})$.
The natural initial conditions would be $U (x, 0)$$=$$0$;
however, as will be shown below, the limiting shape of the foam does not
depend on the initial conditions if $\eta(t)$ is ergodic and $k$$=$$0$.

In (\ref{eq1})  move $U (x, t)$ to the left-hand side, divide both parts of by $\Delta t$, and take the
limit as $\Delta t \rightarrow 0$, to obtain 
\begin{equation}
\label{eq2}
  \frac{\partial U (x, t)}{\partial t} = - g(x - \eta) - k U (x, t).
\end{equation}
It can be shown by numerical simulation with some arbitrary $\eta (t)$, that the solution $U (x, t)$ has
a linear trend, i.e. it behaves as a linearly decaying function of $t$ with superimposed 
fluctuations. We wish to eliminate this trend and see if we can achieve some sort of stationary behavior of $U (x, t)$. Perform the change of variables
\begin{eqnarray*}
V = \frac{U}{t}, \quad \frac{\partial V}{\partial t} =  \frac{1}{t} \left(
   \frac{\partial U}{\partial t} - V \right), \quad 
  \frac{\partial U}{\partial t} = t \frac{\partial V}{\partial t} + V, 
\end{eqnarray*}
and rewrite (\ref{eq2}) as follows
\begin{equation}
\label{eq3}
  \frac{\partial V}{\partial t} = - \frac{1}{t} \bigg( V + g(x - \eta)  \bigg) - k V.
\end{equation}
%Evolution of the foam profile $V(x,t)$ is illustrated in Fig. {\ref{fig_uncor_cor}}: (a) in 3D,  and (b) in its projection on the $(x,t)$ plane, as the signal shown by filled circles in (b) is applied at each consecutive time moment $t$. 
Eq. (\ref{eq3}) has the same form if the  stimulus $\eta$ is a vector  of  dimension $N$; then $x$ is a vector, and $V$ and $g$ are functions of $N$ variables. 
\begin{figure}
\includegraphics[width=0.5\textwidth]{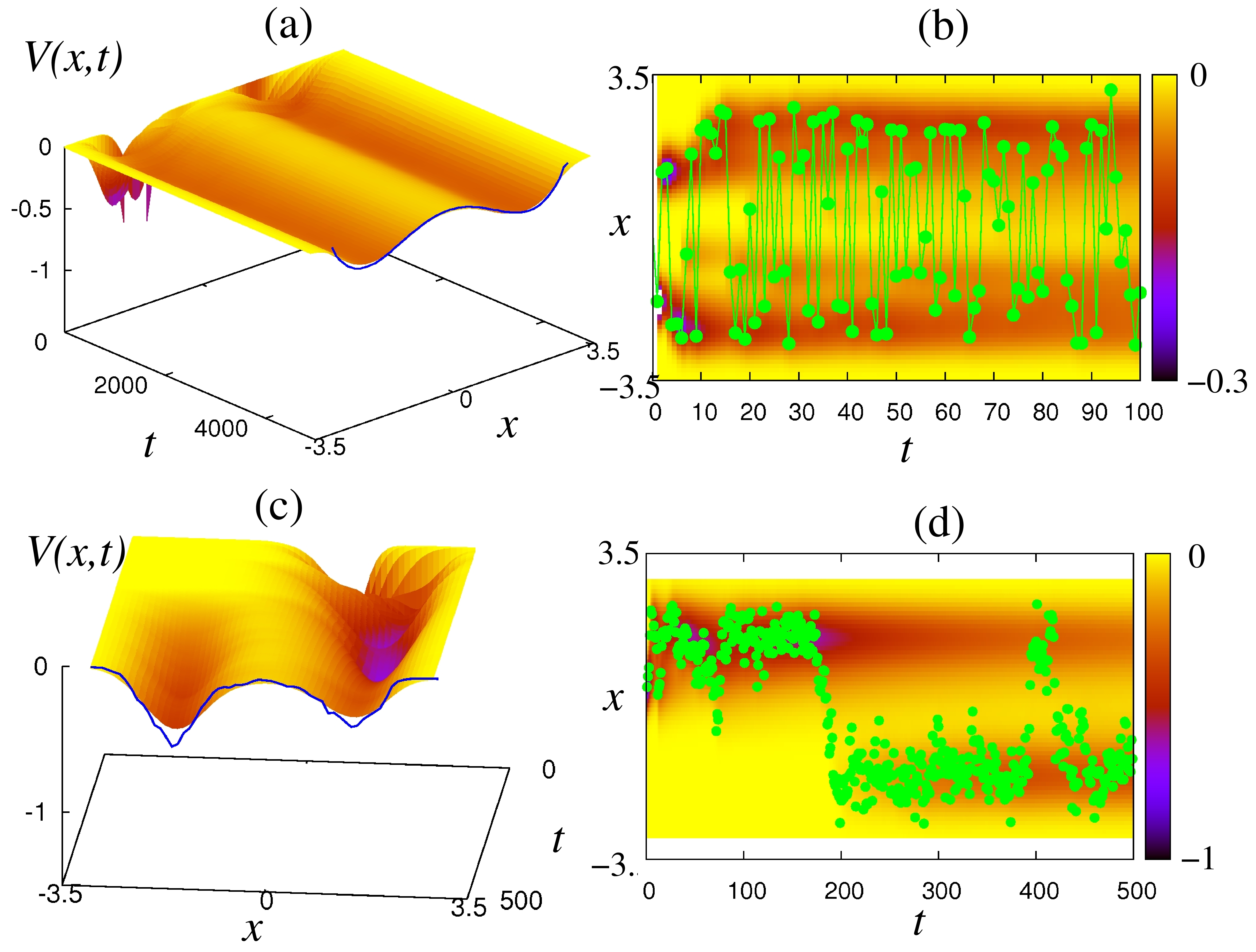}
\caption{\label{fig_uncor_cor}  (Color online) Evolution of the ``memory foam" $V(x,t)$ as the random stimulus is applied by numerically simulating Eq. (\ref{eq3}): (a,c) 3D view; (b,d) projection of $V(x,t)$ onto $(x,t)$ plane shown by color (shade of grey), and the stimulus applied -- by filled circles. In (a,c) the probability density distribution of stimulus is given by solid line at the front. 
In (a,b) the consecutive values of the stimulus are uncorrelated, and in (c,d) -- correlated \cite{comment2}. }
\end{figure}

~\vspace{-2mm}

In Fig. \ref{fig_uncor_cor} the evolution of $V(x,t)$ is illustrated, as two kinds of scalar stimuli are applied to the one-dimensional  foam: 
%Evolution of the foam profile $V(x,t)$ is illustrated in Fig. {\ref{fig_uncor_cor}}: 
(a), (c) show the evolving $V$ as a surface in 3D,  and (b), (d) show $V$ by the color on the $(x,t)$ plane, as the signal shown by filled circles is applied at each consecutive time moment $t$. The PDDs of the two stimuli are of similar two-peak shape (see solid lines at the front in (a,c)), but their two consecutive values are non-correlated in (a,b), and correlated in (c,d)  \cite{comment2}. 
%The actual signals applied are shown by filled circles in (b,d), and 
For both simulations in $g(z)$ we used $\sigma_z$$=$$\sqrt{0.1}$. One can see that eventually both foams shape into the respective PDDs, but if the stimulus values are uncorrelated, the convergence is faster. 

This shaping mechanism reminds the one of the kernel density estimation used in statistics  \cite{Scott_kernel_92}, but is dynamical as opposed to algorithmic, and has no restriction of independent inputs to the system.
If $H(t)$ is not stationary, the foam evolves into a time-averaged density of the input.

\begin{figure}
%\scalebox{0.8}{\includegraphics{fig_flute_1d.eps}}
\includegraphics[width=0.4\textwidth]{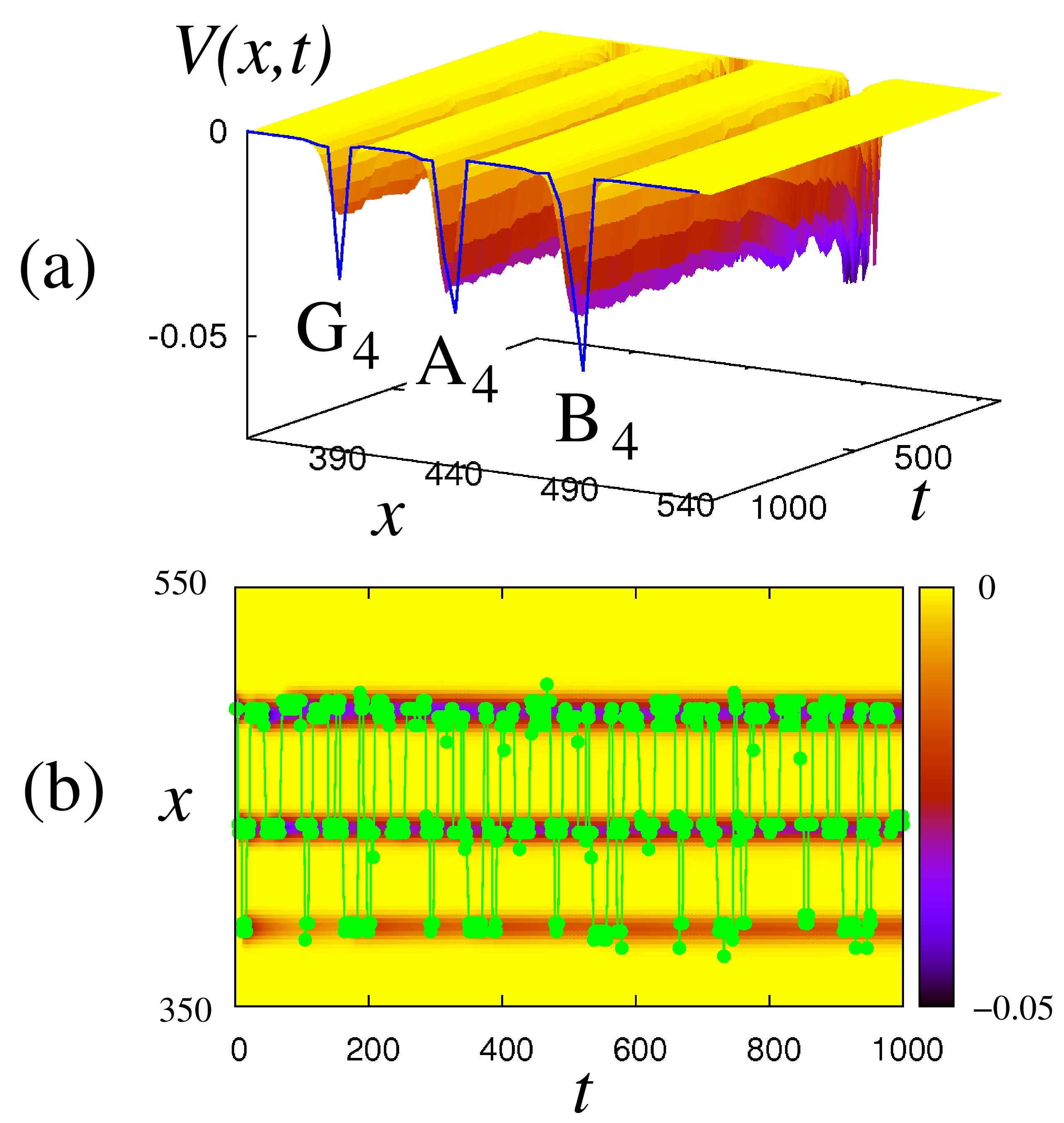}
\caption{\label{fig_flute_1d} (Color online.) Flute -- musical note recognition. Notations are as in Fig.~\ref{fig_uncor_cor}. }
\end{figure}

~\vspace{-2mm}

\noindent {\bf Application to musical data.} 
Next, we illustrate how the proposed foam discovers and memorises musical notes and phrases. 
A children's song ``Mary had a little lamb" was performed with a flute by an amateur musician six times.  The song involves three musical notes ($A$, $B$ and $G$), consists of 32 beats and was chosen for its simplicity to illustrate the principle. The signal was recorded as a 
wave-file  with sampling rate $8$kHz. In agreement with what is usually done in speech recognition \cite{Flanagan_77}, 
the short-time Fourier Transform was applied \cite{Allen_SFT} to the waveform with a sliding window of duration $\tau$$=$$0.75$ sec, which was roughly the duration of each note. The highest spectral peak was extracted for each window, which corresponded to the main frequency $f$ Hz of the given note. A sequence of frequencies $f(t)$ was used to stimulate the foam. 
Note, that each value of $f(t)$ was slightly different from the exact frequency of the respective note, because of the natural variability introduced by a human musician, and the signal $f(t)$ was in fact random, as seen from Fig. \ref{fig_flute_1d}(b).

First, we illustrate how individual musical notes can be automatically identified. A one-dimensional foam received the signal $\eta(t)$$=$$f(t)$, resampled to $8$Hz to save computation time.  Function $f(t)$ can be seen as a realization of a 1st-order stationary and ergodic  process  $F(t)$, consisting of infinitely many repetitions of the same song, which we observe during finite time. This process has a one-dimensional PDD $p_1^F(f)$, which does not change in time.
Gaussian kernel $g(z)$ was used with  $\sigma_z$$=$$\sqrt{5}$ Hz. As shown in 
Fig.~\ref{fig_flute_1d}(a), the foam converges to some PDD shown by solid line. It 
automatically discovers the most probable frequencies as follows,  figures in brackets showing the exact frequencies of the respective musical notes: 
434Hz (440Hz) for $A_4$, 490Hz (493.88Hz) for $B_4$, and 388Hz (392Hz) for $G_4$. 
\begin{figure}
%\scalebox{0.9}{\includegraphics{fig_flute_4d_ill.eps}}
\includegraphics[width=0.5\textwidth]{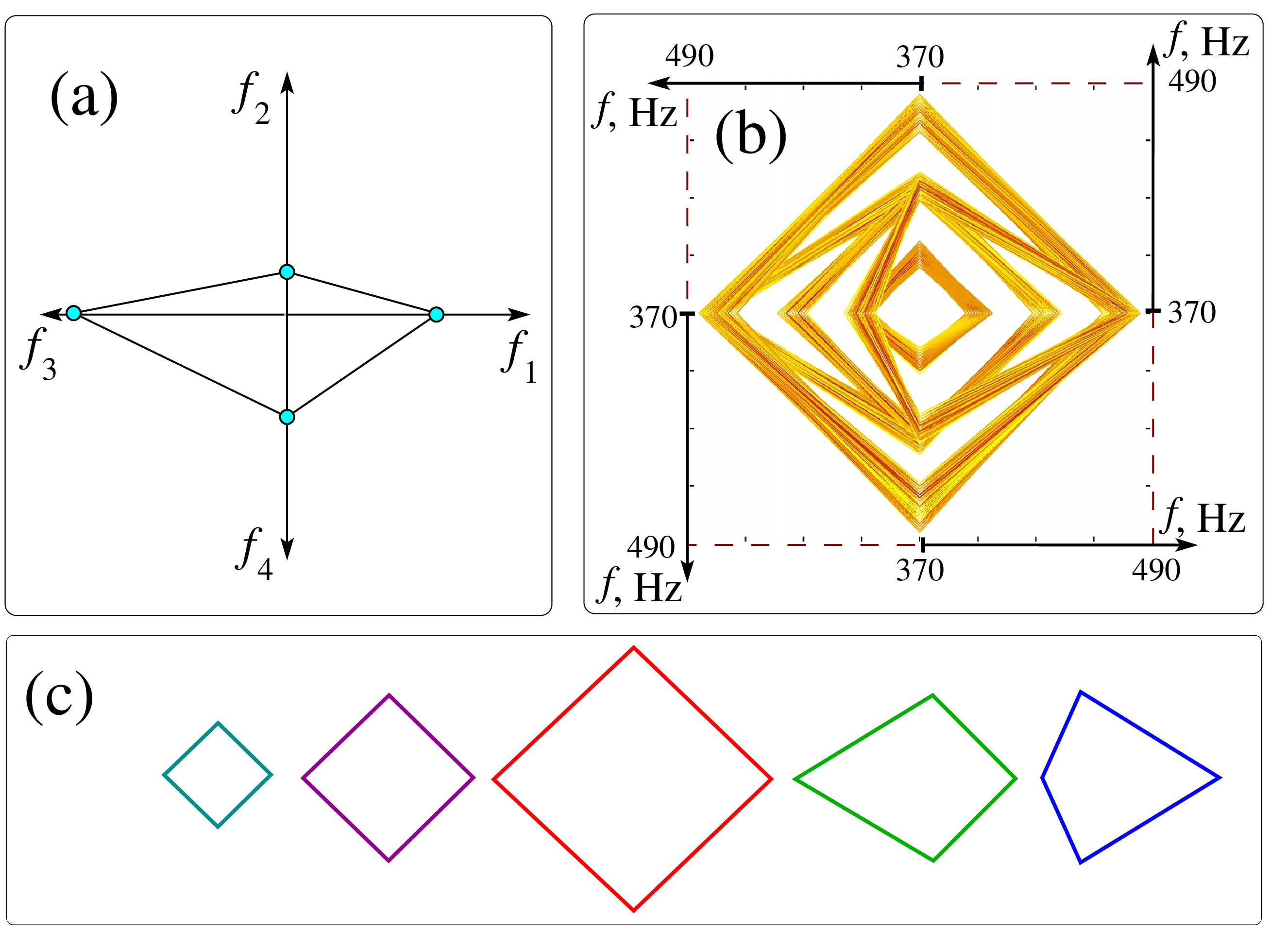}
\caption{\label{fig_flute_4d} (Color online.)  Musical phrase recognition. Description is in text.}
\end{figure}

Second, we show how the foam can discover and memorize temporal patterns -- musical phrases consisting of four beats. The 4D foam was used, and to each of its channels the same signal $f(t)$ was applied, but with a phase shift. Namely, at each time $t$ the foam received a vector stimulus $\psi(t)$$=$$(f(t),f(t+\tau),f(t+2\tau),f(t+3\tau))$, $\tau$$=$$0.75$ sec. For the purpose of this part, we can regard  $\psi(t)$ as a realization of a 4th-order stationary and ergodic vector random process $\Psi (t)$ (which we observe during finite time) with $4$D PDD $p^{\Psi}_4(f_1,f_2,f_3,f_4)$. 
We used a multivariate Gaussian kernel $g$ with $\sigma_z$$=$$\sqrt{5}$ Hz in all of its four variables. 

One cannot visualize evolution of a 4D foam in the same way as we did in Figs.~\ref{fig_uncor_cor}-\ref{fig_flute_1d}, and we
use an alternative representation. We take four half-axes and make their origins coincide (Fig.~\ref{fig_flute_4d}(a)). 
For each feasible input $\psi$$=$$(f_1,f_2,f_3,f_4)$ we put 4 points with coordinates $f_i$ on each of half-axes, and connect them by lines. Thus, any feasible input pattern is represented by a polygon on a plane. (This can be done for any dimension of input vector.) The value of $p^{\Psi}_4$ at each point can be represented by the color of the respective polygon (Fig.~\ref{fig_flute_4d}(b)). The polygon, whose color is the darkest, is the most probable pattern. Unfortunately, when too many polygons overlap, it might be difficult to see the darkest ones. But they can be found using a particle in the 4D foam, that will go to the most probable pattern: five such patterns are given in smaller scale in Fig.~\ref{fig_flute_4d}(c). 

Recognition of musical phrases is also illustrated by the supplemented wave-files \cite{URL}.

~\vspace{-2mm}

\noindent {\bf Discussion.} The memory foam approach presented here could pave the way to create a new generation of information processing machines. 
Unlike digital computers or discrete-state neural networks, 
these devices will be fully analogue and in this sense closer to biological brains.
The proposed approach assumes naturally unsupervised learning, which is traditionally more challenging than other types of learning; however, supervision can be  implemented at any stage, if required. Also, the ``memory foam" can combine learning with pattern recognition, i.e. function in the ``on-line learning" regime. 
With a musical example we demonstrated how hierarchies of patterns can be created in a dynamical way, by going from single notes to their combinations. The importance of being able to create hierarchies of patterns in AI devices cannot be overestimated (see e.g. \cite{Hawkins_Intelligence}). 

%A famous major problem, arising in connection with AI performance, is the so-called ``curse of dimensionality". As the problem becomes more complicated,  the number of states of a traditional AI device grows very quickly, and becomes too large for the computer memory, or the connectivity of artificial NNs. The ``curse" can be worked around \cite{Powell}, but there is always a price (e.g. the duration of calculations). The ``memory foam" device would not require connectivity similar to that in NNs, and might provide a solution to the ``curse" problem. 

%~\vspace{-2mm}

\noindent {\bf Acknowledgements.}
The authors are grateful to Alexander~Balanov for a number of helpful critical comments on the draft of this paper, and to Victoria~Marsh for playing the flute. 

%\bibliography{foam}

\begin{thebibliography}{10}
\bibitem{comment1}
One can speak about an energy profile if neurons are coupled symmetrically.
\bibitem{comment2}
The stimulus illustrated in Fig. \ref{fig_uncor_cor} (a,b) is obtained by
  taking Gaussian white noise and applying a non-linear transformation, that
  changed its probability density distribution (PDD). Thus, the PDD took the
  shape shown in (a) by solid line, but the consecutive values remained
  uncorrelated. The stimulus in (c,d) is obtained by applying Gaussian white
  noise to a differential equation describing a particle moving in a
  non-symmetric double-well potential with large viscosity \cite{Malakhov_97}.
  The PDD of the output signal has the shape shown in (c) by solid line, and
  the consecutive values are correlated.
\end{thebibliography}
%\bibliographystyle{phaip}

\end{document}